# Spatially-coded Fourier ptychography: flexible and detachable coded thin films for quantitative phase imaging with uniform phase transfer characteristics

Ruihai Wang[1,†], Liming Yang[1,†], Yujin Lee[2,†], Kevin Sun[3], Kuangyu Shen[4], Qianhao Zhao[1], Tianbo Wang[1], Xincheng Zhang[1], Jiayi Liu[5], Pengming Song[1,*] and Guoan Zheng[1,*]

[1]Department of Biomedical Engineering, University of Connecticut, Storrs, USA
[2]School of Electrical and Electronic Engineering, Yonsei University, Seoul, Republic of Korea
[3]Andover High School, Andover, USA
[4]Department of Chemical and Biomolecular Engineering, University of Connecticut, Storrs, USA
[5]Farmington High School, Farmington, USA
[†]These authors contributed equally to this work
[*]Correspondence: pengming.song@uconn.edu (P.S.); guoan.zheng@uconn.edu (G.Z.)

**Abstract**: Fourier ptychography (FP) is an enabling imaging technique that produces high-resolution complex-valued images with extended field coverages. However, when FP images a phase object with any specific spatial frequency, the captured images contain only constant values, rendering the recovery of the corresponding linear phase ramp impossible. This challenge is not unique to FP but also affects other common microscopy techniques -- a rather counterintuitive outcome given their widespread use in phase imaging. The underlying issue originates from the non-uniform phase transfer characteristic inherent in microscope systems, which impedes the conversion of object wavefields into discernible intensity variations. To address this challenge, we present spatially-coded Fourier ptychography (scFP), a new method that synergizes FP with spatial-domain coded detection for true quantitative phase imaging. In scFP, a flexible and detachable coded thin film is attached atop the image sensor in a regular FP setup. The spatial modulation of this thin film ensures a uniform phase response across the entire synthetic bandwidth. It improves reconstruction quality and corrects refractive index underestimation issues prevalent in conventional FP and related tomographic implementations. The inclusion of the coded thin film further adds a new dimension of measurement diversity in the spatial domain. The development of scFP is expected to catalyse new research directions and applications for phase imaging, emphasizing the need for true quantitative accuracy with uniform frequency response.

**Date of submission**: November 26, 2023

The quest for higher resolution and information-rich imaging has been a driving force in the development of advanced optical microscopy techniques[1]. Fourier ptychography (FP) is one example that tackles the intrinsic trade-off between resolution and field of view in imaging systems[2]. It allows researchers to achieve high resolution and extended field coverage without involving mechanical scanning. By integrating traditional microscope hardware with advanced computational techniques, FP transforms the general challenge of high-throughput and high-resolution imaging from one that is coupled to the physical limitations of optics to one that is solvable through computation[3].

The FP concept is rooted in the integration of synthetic aperture[4] and phase retrieval[5,6] for non-interferometric super-resolution imaging. A typical FP setup can be built using a regular light microscope with a programmable LED array. In its operation, the LED array sequentially illuminates the specimen from different incident angles and the FP system records the corresponding low-resolution intensity images using a low numerical aperture (NA) objective lens. If the specimen is a 2D thin section, illuminating it with a tilted planewave is equivalent to shifting the object spectrum in the Fourier domain. Therefore, each captured image corresponds to the object information from a distinct circular aperture in Fourier space. The size of the aperture is determined by the NA of the objective lens while its offset from the origin is determined by the illumination wavevector. Unlike methods such as holography that directly measure phase from interferometric measurements, FP recovers the phase information in a post-measurement recovery process. This is often achieved through an iterative algorithm that alternates between spatial and Fourier domains, where constraints can be applied from the captured images and the confined pupil aperture support. Once the object spectrum is recovered in Fourier space, the object image with both intensity and phase properties can be obtained by transforming the synthesized spectrum back to the spatial domain. The final achieved resolution of the recovered object image is no longer limited by the NA of the employed objective lens. Instead, it is determined by the maximum incident angle of the LED array. Meanwhile, the recovered image retains the original large field of view of the employed low-NA objective lens. Furthermore, the rich information provided by an FP dataset not only aids in reconstructing the complex-valued object image but also enables the computational correction of pupil aberrations and other system imperfections post-measurement[7-11].

As its name suggests, FP is intricately linked to an imaging approach termed ptychography. Conventional ptychography is a lensless technique originally developed for addressing the missing phase problem in electron microscopy[12]. Its modern form[13] involves translating the object across a confined illumination probe beam in real space while capturing the resultant diffraction patterns in reciprocal space, all without the use of lenses. The lensless nature of this technique is particularly appealing to researchers in the extreme ultraviolet and X-ray fields[14,15], where fabricating high-resolution lenses is not only expensive but also presents significant technical challenges.

Compared with lensless ptychography, FP employs a lens-based microscope to swap the roles of real and reciprocal spaces. In lensless ptychography, the confined probe beam provides the finite support constraint in real space, whereas in FP, this constraint is applied in reciprocal space by the confined pupil aperture of the objective lens[2,16]. Both techniques share a similar imaging model that necessitate overlapping apertures between adjacent acquisitions to resolve phase ambiguity[17]. However, the lens elements in FP can correct for chromatic dispersion across various wavelengths, leading to less stringent requirements on the temporal coherence of the light source[3]. Consequently, broad-band LED sources are well-suited for sample illumination in FP, while laser sources, with their superior coherence properties, are generally preferred in lensless ptychography[17].

Since the first demonstration in 2013, FP has evolved into a versatile technique for different research communities. This journey has been marked by several unexpected developments of its variants. Notable among these are the creation of a long-range synthetic aperture imaging scheme that deviates from the original microscope configuration[18-20], the realization of an X-ray FP variant for its superior dose efficiency[21], and the innovative integration with diffraction tomography for high-resolution 3D imaging[22-24]. Additionally, the simplicity of an FP setup renders it an excellent education tool in undergraduate laboratories for instructing students in Fourier optics[25].

Despite its rapid advancement and growing popularity in computational microscopy, FP's effectiveness is often hampered by the loss of phase information in its reconstructions. This issue arises from the non-uniform phase transfer characteristic inherent in microscopy systems. To illustrate this challenge, we consider imaging a specimen with a linear phase ramp using a regular FP setup. The phase of this specimen corresponds to a single spatial frequency in Fourier space. When the FP system illuminates the specimen at varying incident angles, the resulting intensity images only contain constant values: non-zero constant for brightfield images and zero for darkfield images. As a result, these captured images offer no or little insight about the specimen's linear phase ramp. This simple thought experiment suggests that FP cannot recover the phase information for any specific spatial frequency, a rather unexpected finding given the technique's previous demonstrations on phase imaging[26-35].

We also note that the phase imaging challenge discussed above is not unique to FP but also affects other common microscopy techniques, such as transport-of-intensity equation[36], support-constraint phase retrieval[37,38], digital in-line holography[39-41], multi-height and multi-wavelength lensless imaging[42,43], blind ptychography[44-46], and differential phase contrast imaging[47]. In FP and these microscopy modalities, despite claims of quantitative phase imaging capabilities, the captured intensity images only contain constant values for any given linear phase ramp, making them difficult to recover the correct phase ramp from intensity measurements. More generally speaking, the capability of capturing phase information can be characterized by phase transfer function (PTF)[36], which measures the transfer of phase content at different spatial frequencies in Fourier space. In the aforementioned modalities, the PTR is zero for any given spatial frequency of phase or a combination thereof. This results in the permanent loss of related phase contents during data acquisition, rendering it unrecoverable post-measurement.

To address this challenge, we present spatially-coded Fourier ptychography (scFP), a new method that synergizes FP with spatial-domain coded detection for true quantitative phase imaging. In scFP, a flexible and detachable coded thin film is attached atop the image sensor in a regular FP setup. The microparticles embedded in the coded thin film effectively convert the object phase into intensity variations, ensuring a uniform frequency response across the entire synthetic bandwidth. This method not only improves ptychographic reconstruction quality but also corrects refractive index underestimation issues prevalent in conventional FP and related diffraction tomographic approaches. The incorporation of the coded thin film further adds a new dimension of measurement diversity to FP, where additional object information can be obtained by translating the specimen or the camera system. The flexible and detachable nature of the film also allows for easy adaptation to different lens-based or lensless microscope systems, establishing it as a versatile tool for a wide range of imaging settings. We validate scFP's efficacy by imaging various samples that are challenging for other common phase imaging techniques. The development of scFP is expected to catalyse new research directions and applications, particularly in the realm of computational microscopy where quantitative accuracy, adaptability, and versatility are of important considerations.

## Results
### Principle of spatially-coded Fourier ptychography (scFP)
The principle of scFP is rooted in the integration of FP with the concept of coded detection for tackling the phase recovery challenge inherent in microscope systems. Figure 1 shows a comparative view of conventional FP alongside the proposed scFP approach. As shown in Figure 1(a), conventional FP illuminates the specimen with different incident angles and captures the corresponding intensity images with a regular image sensor. The object specimen in Figure 1 is a phase-only target with half 0 and half $\pi$ phase, with a step response to better reveal the PTF of the imaging modalities. The right panel of Figure 1(a) shows the simulated raw image and the recovered phase using conventional FP. In this example, only the edge between 0 and $\pi$ phase can be converted into intensity variations in the raw image. As a result, conventional FP fails to recover the correct phase values of the target, with all low-frequency phase contents missing in the recovered phase. This limitation underscores the challenge in accurately retrieving phase information using conventional FP and other common microscopy techniques.

Figure 1(b) introduces the proposed scFP setup, where a flexible and detachable coded thin film is attached on top of the image sensor in a conventional microscope setup. This coded thin film contains microparticles that spatially encode the complex wavefronts of the object, converting the phase information into detectable intensity variations. The right panels of Figure 1(b) show the simulated raw image and the recovered phase using scFP. In this case, the raw intensity image is a speckle-like pattern that presents the modulation between the input object phase with the coded thin film. As shown in the right panel of Figure 1(b), the recovered phase by scFP converges to the correct phase values as the input ground truth.

By analysing the spatial-frequency contents of the input ground-truth phase and the recovered phase images, we obtain the PTFs in Figures 1(c) and 1(d). For conventional FP, the low-frequency phase contents are notably absent while the high-frequency phase contents are presented thanks to the modulation of pupil aperture of the low-NA objective lens. Importantly, employing a high-NA objective lens during image acquisition ironically results in the omission of this high-frequency phase content. In an idealized imaging system with an infinite NA, the intensity profile of a phase object would remain invariant, rendering phase recovery from intensity measurements unattainable. In stark contrast, the proposed scFP achieves a uniform response across the entire synthetic bandwidth, overcoming the major limitation faced by conventional FP as well as other common microscopy techniques.

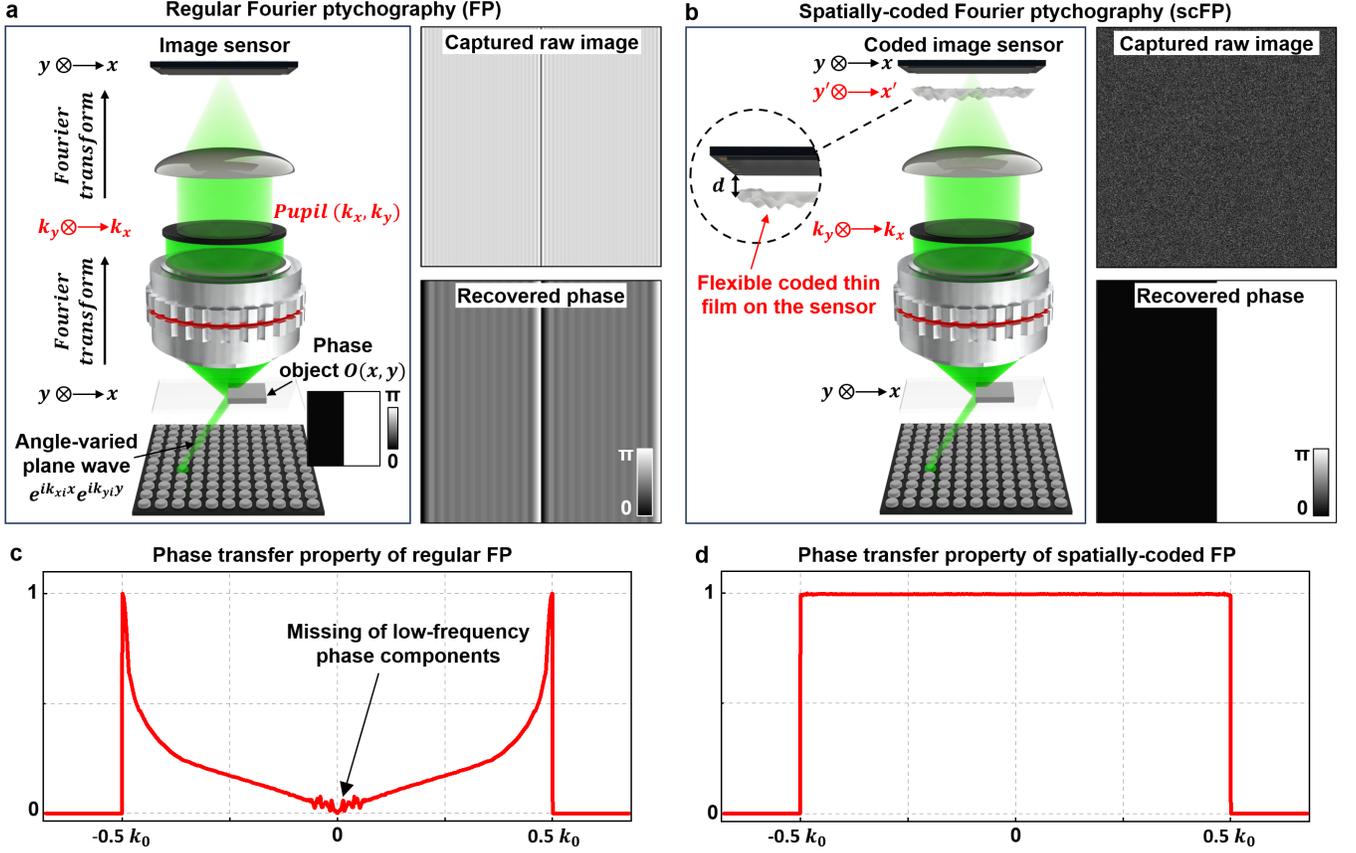

**Figure 1. Principles of conventional FP and spatially-coded FP.** For both approaches, we assume using a 2×, 0.1-NA objective lens for image acquisition and the final synthetic NA is 0.5. The object specimen is phase-only target with half 0 and half $\pi$ phase, with a step response to better reveal the PTF of both approaches. (a) Conventional FP: The specimen is sequentially illuminated with different incident angles and intensity images are acquired using a regular image sensor. The right panel illustrates the simulated raw image and the recovered phase of the phase target, highlighting the limitation in retrieving low-frequency phase contents. (b) The proposed scFP approach with a flexible and detachable coded thin film attached on the image sensor. The right panels show the raw intensity image and the recovered phase of scFP, demonstrating its superior performance in recovering the correct phase. Spatial-frequency analysis of the phase transfer property for conventional FP (c) and scFP (d). For conventional FP, the low-frequency phase contents are missing, while the high-frequency content can be recovered thanks to the modulation of the low-NA objective lens. In contrast, scFP achieves a uniform response across the entire synthetic bandwidth, demonstrating its true quantitative phase imaging capability.

The imaging model of scFP can be explained as follows. The object $O(x, y)$ is illuminated by a planewave $e^{ik_{xi}x + ik_{yi}y}$ from the i$^{th}$ element of the LED array, with the incident wavevector denoted as $(k_{xi}, k_{yi})$. The object spectrum, before modulation by the pupil aperture of the objective lens, is given by:

$$\hat{O}(k_x - k_{xi}, k_y - k_{yi}) = \text{FT}\{O(x, y) \cdot e^{ik_{xi}x + ik_{yi}y}\}, \tag{1}$$

where 'FT' denotes Fourier transform, '·' denotes point-wise multiplication. The object spectrum then passes through the pupil aperture of the objective lens and the resulting wavefield propagates to the plane $(x', y')$ of the coded thin film:

$$W_i(x', y') = \text{Prop}_{-d}\{\text{FT}[\hat{O}(k_x - k_{xi}, k_y - k_{yi}) \cdot \text{Pupil}(k_x, k_y)]\}, \quad (2)$$

where '$\text{Prop}_{-d}$' represents free-space propagation for a distance $-d$, and '$\text{Pupil}(k_x, k_y)$' denotes the pupil aperture of the objective lens. As shown in Figure 1(b), we assume the image sensor is placed at the in-focus position and the coded thin film is placed a distance of the $d$ away from the sensor. The term '$-d$' in Eq. (2) means back-propagation from the plane of the detector $(x, y)$ to the plane of the coded thin film $(x', y')$. The wavefield in Eq. (2) is then modulated by the transmission profile of the coded surface of the thin film, $CS(x', y')$. The i$^{th}$ captured intensity image $I_i(x, y)$ on the detector can be expressed as:

$$I_i(x, y) = |\text{Prop}_d\{W_i(x', y') \cdot CS(x', y')\}|^2 \quad (3)$$

Compared to conventional FP, scFP introduces a back-propagation step '$\text{Prop}_{-d}$' in Eq. (2), an encoding step with $CS(x', y')$ in Eq. (3), and a propagation step '$\text{Prop}_d$' in Eq. (3). With the captured images $I_i(x, y)$, scFP aims to recover the high-resolution complex profile of object $O(x, y)$.

---

Initialization: $O(x, y) = \frac{1}{K}\sum_{i=1}^{K}\sqrt{I_i(x,y)}$, $\hat{O}(k_x, k_y) = \text{FT}\{O(x,y)_{\uparrow(M/m)}\}$, $\text{Pupil}(k_x, k_y) = circ\left(NA \cdot \frac{2\pi}{\lambda}\right)$

1. **for** $n = 1 : N$ **(different iterations)**
2.   **for** $i = 1 : I$ **(different incident angles)**
3.     Crop a subregion from the Fourier spectrum ($k_{xi}, k_{yi}$: the shift in the Fourier domain)

$$\hat{O}_{i\_crop}(1:m, 1:m) = \hat{O}\left(\frac{M-m}{2} + k_{xi} + 1 : \frac{M+m}{2} + k_{xi}, \frac{M-m}{2} + k_{yi} + 1 : \frac{M+m}{2} + k_{yi}\right)$$

4.     $\hat{O}_{i\_fil}(k_x, k_y) = \hat{O}_{i\_crop}(k_x, k_y) \cdot \text{Pupil}(k_x, k_y)$     % Low-pass filtering
5.     $W_i(x, y) = \text{FT}^{-1}\{\hat{O}_{i\_fil}(k_x, k_y)\} * \text{PSF}_{\text{free}}(-d)$     % Propagate the wavefield to the coded surface plane
6.     $E_i(x, y) = W_i(x, y) \cdot CS(x, y)$     % Exit wave at the coded surface plane
7.     $O_{i\_fil}(x, y) = E_i(x, y) * \text{PSF}_{\text{free}}(d)$     % Propagate the exit wave to the sensor plane
8.     $O'_{i\_fil}(x, y) = \sqrt{I_i(x, y)} \cdot \frac{O_{i\_fil}(x,y)}{|O_{i\_fil}(x,y)|}$     % Replace the amplitude and keep the phase unchanged
9.     $E'_i(x, y) = O'_{i\_fil}(x, y) * \text{PSF}_{\text{free}}(-d)$     % Propagate to the coded surface plane
10.    Update the wavefield at the coded surface plane using the rPIE algorithm:

$$W_i^{update}(x, y) = W_i(x, y) + \frac{\text{conj}(CS(x,y)) \cdot \{E'_i(x,y) - E_i(x,y)\}}{(1-\alpha_W)|CS(x,y)|^2 + \alpha_W|CS(x,y)|^2_{max}}$$

11.    $O_{i\_fil}^{update}(x, y) = W_i^{update}(x, y) * \text{PSF}_{\text{free}}(d)$     % Propagate the updated wavefield to the sensor plane
12.    $\hat{O}_{i\_fil}^{update}(k_x, k_y) = \text{FT}\{O_{i\_fil}^{update}(x, y)\}$
13.    Update $\hat{O}_{i\_crop}(k_x, k_y)$ and $\text{Pupil}(k_x, k_y)$ using the rPIE algorithm

$$\hat{O}_{i\_crop}^{update}(k_x, k_y) = \hat{O}_{i\_crop}(k_x, k_y) + \frac{\text{conj}(\text{Pupil}(k_x,k_y)) \cdot \{\hat{O}_{i\_fil}^{update}(k_x,k_y) - \hat{O}_{i\_fil}(k_x,k_y)\}}{(1-\alpha_O)|\text{Pupil}(k_x,k_y)|^2 + \alpha_O|\text{Pupil}(k_x,k_y)|^2_{max}}$$

$$\text{Pupil}^{update}(k_x, k_y) = \text{Pupil}(k_x, k_y) + \frac{\text{conj}(\hat{O}_{i\_crop}(k_x,k_y)) \cdot \{\hat{O}_{i\_fil}^{update}(k_x,k_y) - \hat{O}_{i\_fil}(k_x,k_y)\}}{(1-\alpha_P)|\hat{O}_{i\_crop}(k_x,k_y)|^2 + \alpha_P|\hat{O}_{i\_crop}(k_x,k_y)|^2_{max}}$$

14.    Update the corresponding subregion of the object spectrum

$$\hat{O}\left(\frac{M-m}{2} + k_{xi} + 1 : \frac{M+m}{2} + k_{xi}, \frac{M-m}{2} + k_{yi} + 1 : \frac{M+m}{2} + k_{yi}\right) = \hat{O}_{i\_crop}^{update}(k_x, k_y)$$

15.   **end**
16. **end**
17.    Perform an inverse Fourier transform of the object spectrum: $O(x, y) = \text{FT}^{-1}\{\hat{O}(k_x, k_y)\}$

**Figure 2. Reconstruction procedures of scFP.** Following the initialization of the object spectrum and pupil aperture, the algorithm undergoes iterative steps to progressively recover the object spectrum. The final step involves transforming the recovered spectrum back to the spatial domain, obtaining the high-resolution object profile.

The reconstruction process of scFP is shown in Figure 2. Starting with the initialization of the object spectrum and pupil aperture, the algorithm iterates through successive refinements of the object spectrum. The final step encompasses the transformation of the spectrum back to the spatial domain, obtaining the high-resolution object profile. With the imaging model and the reconstruction algorithm, we performed a series of simulation in Figure 3 to evaluate the performance of scFP, particularly in the reconstruction of phase objects. Figure 3(a) shows the input ground truth intensity and phase for the simulation study. Figures 3(b) and 3(c) show the recovered images using conventional FP and scFP, respectively. We can see that scFP demonstrates a clear and accurate reconstruction of the phase ramp, mirroring the ground truth. Further examination through line profile analysis in Figure 3(d) quantifies the improvement scFP brings to phase imaging. The line profiles extracted from the scFP reconstructions align closely with the ground truth, while conventional FP exhibits notable discrepancies.

In Figure 1(c) and 1(d), we show the PTFs for the phase target with a step function. An exact PTF of a specimen allows for the potential of deconvolution to refine the phase image recovery. However, the coupling of PTF with the object's intensity profile complicates this process; differing objects yield variant PTFs, rendering deconvolution infeasible due to the indeterminate nature of the object's properties. To clarify this challenge, we conducted a simulation employing a constant intensity input alongside a linear phase ramp, as shown in Figure 3(e). The recovered phase by conventional FP is different from that in Figure 3(d), despite identical phase ramps. Conversely, scFP consistently recovers accurate phase information, unaffected by the object's intensity variations, thus affirming its status as an effective tool for true quantitative phase imaging with a uniform frequency response.

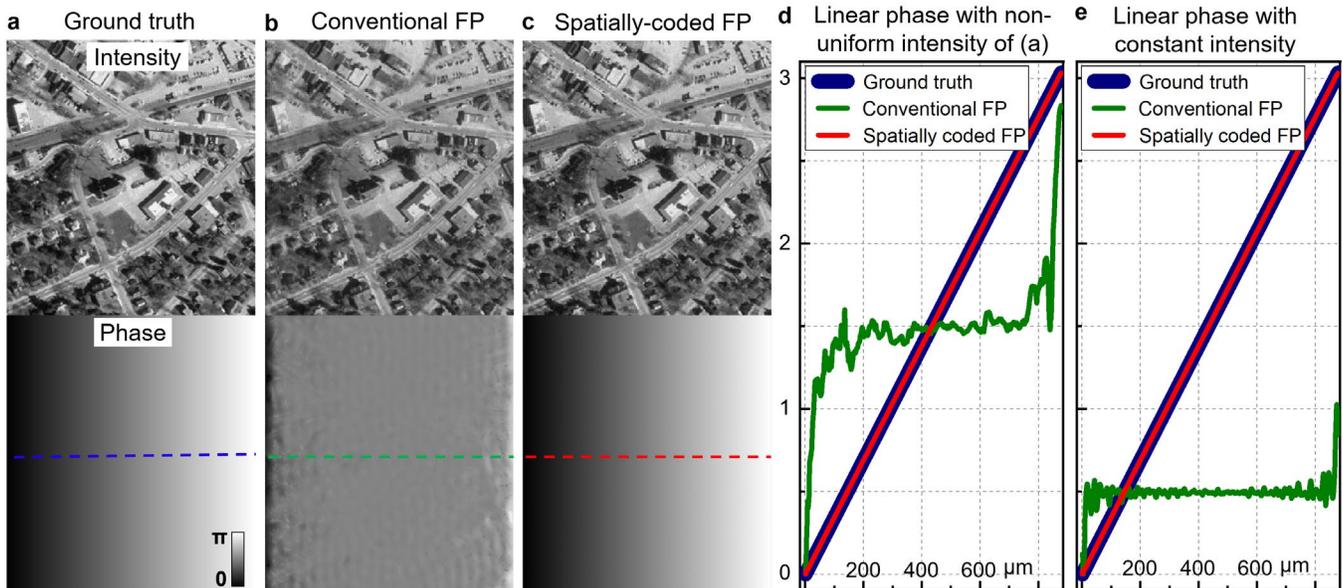

**Figure 3. Comparative simulations of conventional FP and scFP.** (a) Grund-truth intensity and phase images for simulation. The recovered intensity and phase using conventional FP (b), and scFP (c). (d) Line traces extracted from the linear phase ramp, comparing the ground truth, conventional FP, and scFP. (e) Simulation conducted for a phase-only object with a linear phase ramp and a constant intensity. scFP exhibits robust and accurate phase reconstruction in alignment with the ground truth, whereas conventional FP displays significant deviation.

**Experimental demonstrations**

The fabrication of the coded thin film is critical for the effectiveness of scFP, as it directly influences the quality of ptychographic reconstruction. The process begins with the preparation of a polydimethylsiloxane (PDMS) mixture, which is thoroughly mixed with microparticles to form the foundation of the thin film (Figure 4(a)). We typically choose polystyrene beads or metal microparticles approximately 1 μm in diameter. The mixture is then evenly spread across a thin film substrate that is secured on a vacuum pad to ensure stability and flatness during the coating process (Figure 4(b)). An adjustable film applicator is utilized to coat the substrate with the PDMS mixture, achieving a consistent layer approximately 2 μm thick (Figure 4(c)). After the base layer solidifies, a layer of clear PDMS is applied on top to increase the film's thickness, facilitating easier handling and removal from the substrate (Figure 4(d)-4(e)). The coated film is subsequently cured to solidify the structure (Figure 4(f)). Upon completion of the curing process, the film is gently peeled from the substrate, yielding a flexible coded thin film that is ready for integration into different image sensors (Figure 4(g) and 4(h)). The resulting flexible and detachable coded film, thin yet densely packed with microparticles, is adept at modulating incident light waves, a property that renders it suitable for use not only in scFP but also in assorted diffraction tomographic and lensless imaging setups.

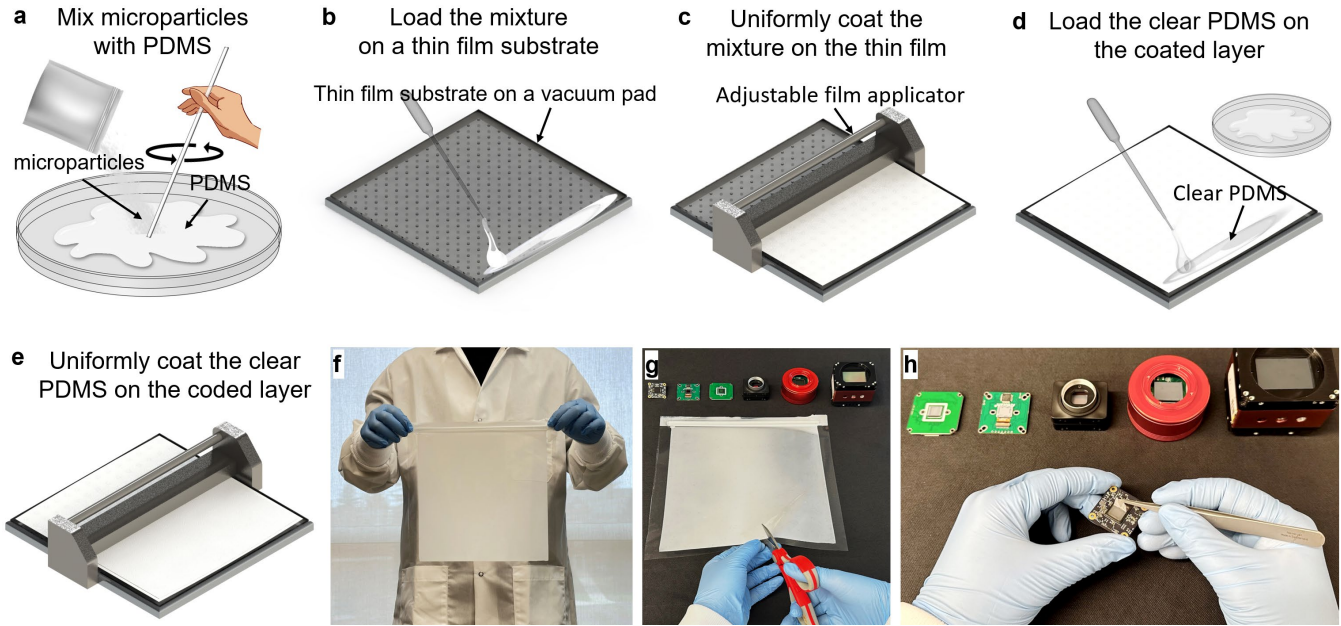

**Figure 4. Fabrication process of the flexible and detachable coded thin film.** (a) A homogeneous mixture of microparticles with PDMS is prepared as the base layer of the coded thin film. (b) The mixture is evenly spread on a plastic film substrate using a vacuum pad to ensure flatness. (c) An adjustable film applicator is used to uniformly coat the mixture, controlling the thickness of the film. (d) A clear layer of PDMS is then loaded over the coded base layer to increase the thickness. (e) The clear PDMS is uniformly spread on top of the coded pattern. (f) The coated film is held up for inspection to ensure an even coating. (g) The film undergoes a curing process and is then carefully cut and peeled from the thin plastic substrate. (h) Integrate the flexible and detachable coded film with different image sensors.

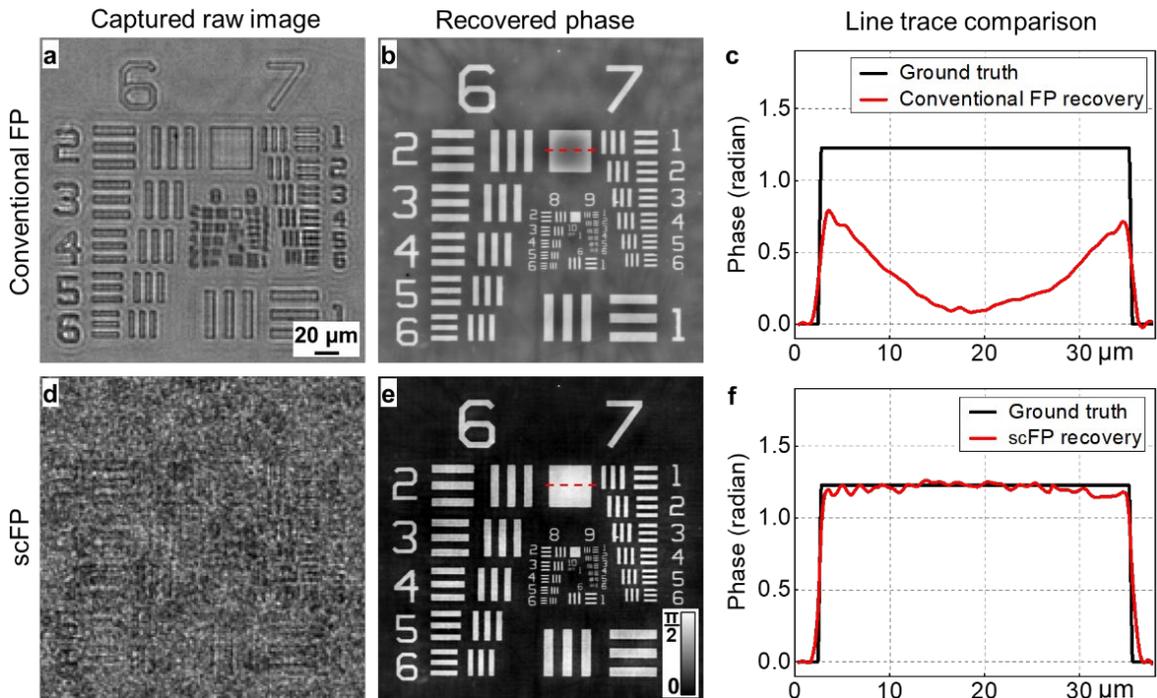

**Figure 5. Validation of scFP using a quantitative phase target**. Captured raw image (a), recovered phase (b), line traces (c) of conventional FP. Captured raw image (d), recovered phase (e), and line traces (f) of scFP. The comparison underscores scFP's proficiency in recovering the complete phase contents, contrasting with the limitations evident in conventional FP.

Figure 5 showcases the accuracy of scFP in reconstructing phase information on a quantitative phase target, underscoring its superior performance over conventional FP. In Figure 5(a), the raw intensity image captured with conventional FP is presented, while Figure 5(b) illustrates the phase image recovered from this data, with the line traces across a square feature detailed in Figure 5(c). These images, particularly Figure 5(c), highlight the inability of conventional FP to capture low-frequency phase content, as evidenced by the line profile's deviation from the ground truth. Conversely, Figures 5(d)-5(f) display the scFP results, with Figure 5(d) revealing a speckle-like pattern indicative of the interaction between the coded thin film and the incident light, a fundamental aspect of scFP's operation. The scFP-recovered phase image in Figure 5(e) closely emulates the ground truth, a relationship that is echoed in the precise line profile of Figure 5(f). The agreement with the ground truth underscores scFP's capacity for true quantitative phase imaging. The experiment cements scFP's status as a potent technique for the accurate reconstruction of phase, particularly beneficial for imaging samples with flat surfaces.

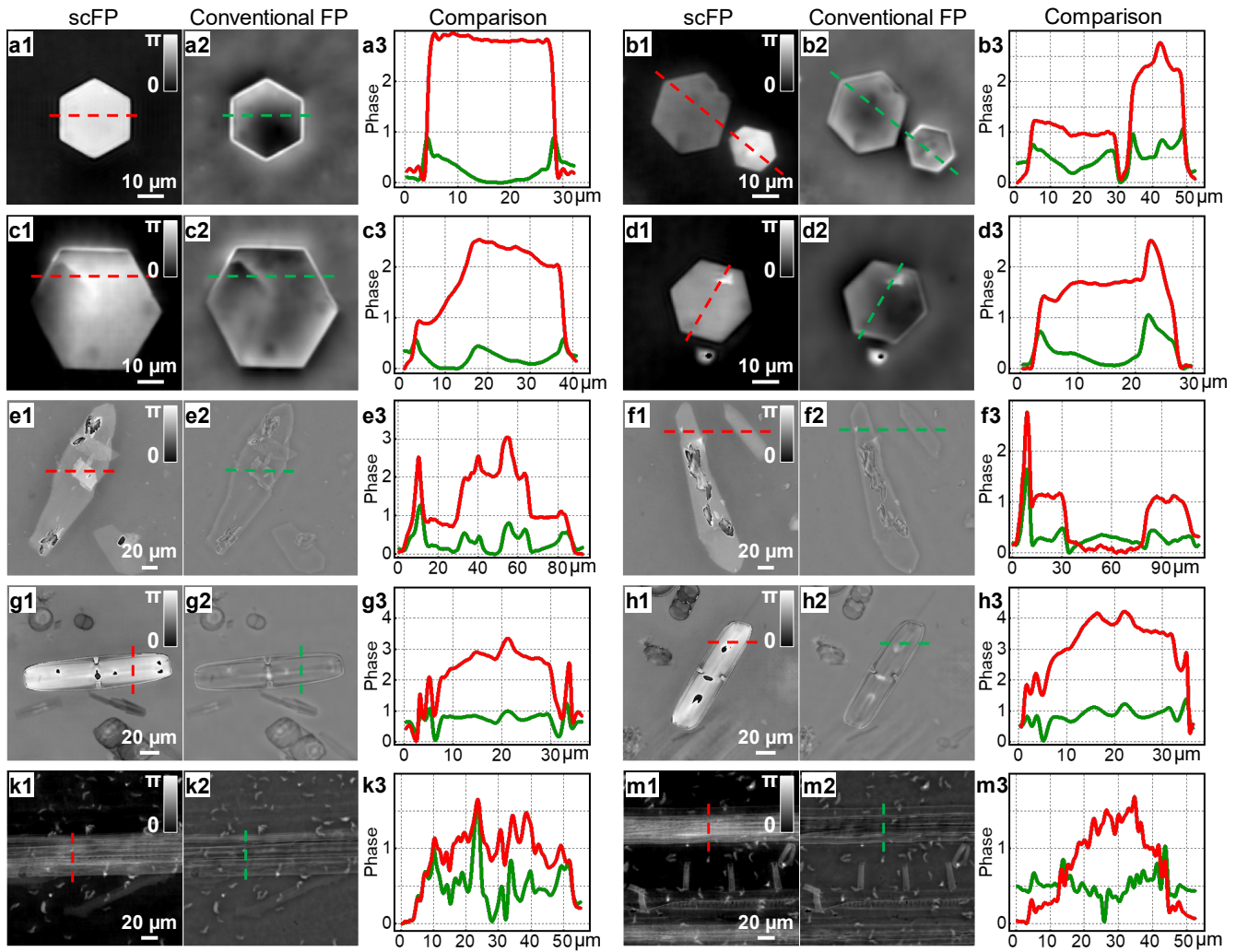

**Figure 6. Validation of scFP across various samples: (a)-(d) cystine crystals, (e)-(f) calcium phosphate crystals, (g)-(h) diatom, and (k)-(m) corn stem slides.** The recovered phase images by scFP (a1-m1), and conventional FP (a2-m2). The line trace comparisons (a3-m3) highlight the scFP recoveries (in red) against the conventional FP (in green), demonstrating the latter's deficiency in capturing low-frequency phase content, a discrepancy effectively resolved by scFP.

Figures 6 and 7 collectively demonstrate the efficacy and adaptability of scFP across an array of sample types, affirming its utility in quantitative phase imaging. Figure 6 displays a variety of phase targets, from crystalline structures to plant tissues. The phase images recovered by scFP and those by conventional FP are presented side-by-side for comparison. Line trace analyses accentuate the scFP's ability to recover the complete phase information, particularly the low-frequency components that conventional FP fails to capture. In Figure 7, scFP's application extends to biological specimens of mouse kidney tissues and Hela cell cultures. The raw and recovered phase images using conventional FP are noticeably less when compared to the scFP counterparts, which display enhanced structural clarity. This capability illustrates scFP's potential as a true quantitative tool in the fields of life sciences and medical diagnostics, providing clearer visibility and better measurement accuracy into the microscopic landscape.

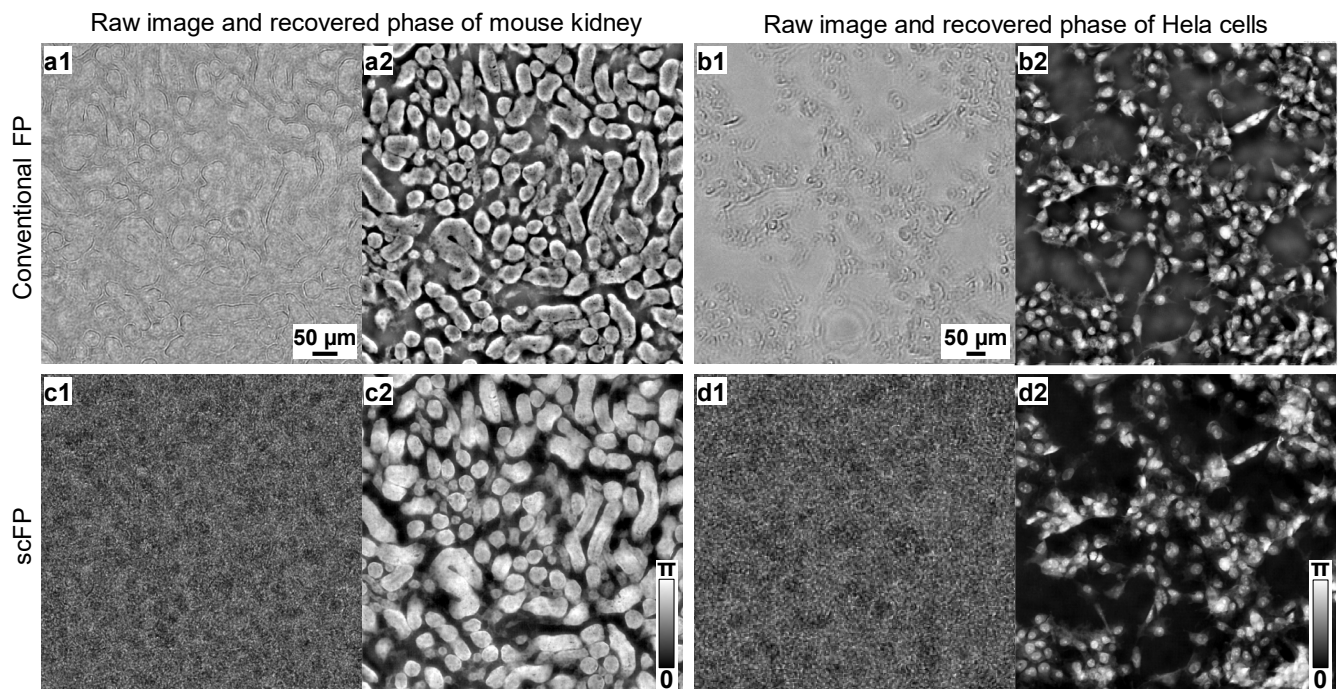

**Figure 7. Validation of scFP on mouse kidney slide and Hela cell cultures**. (a)-(b) Raw images and recovered phase using conventional FP. (c)-(d) Raw images and recovered phase using scFP.

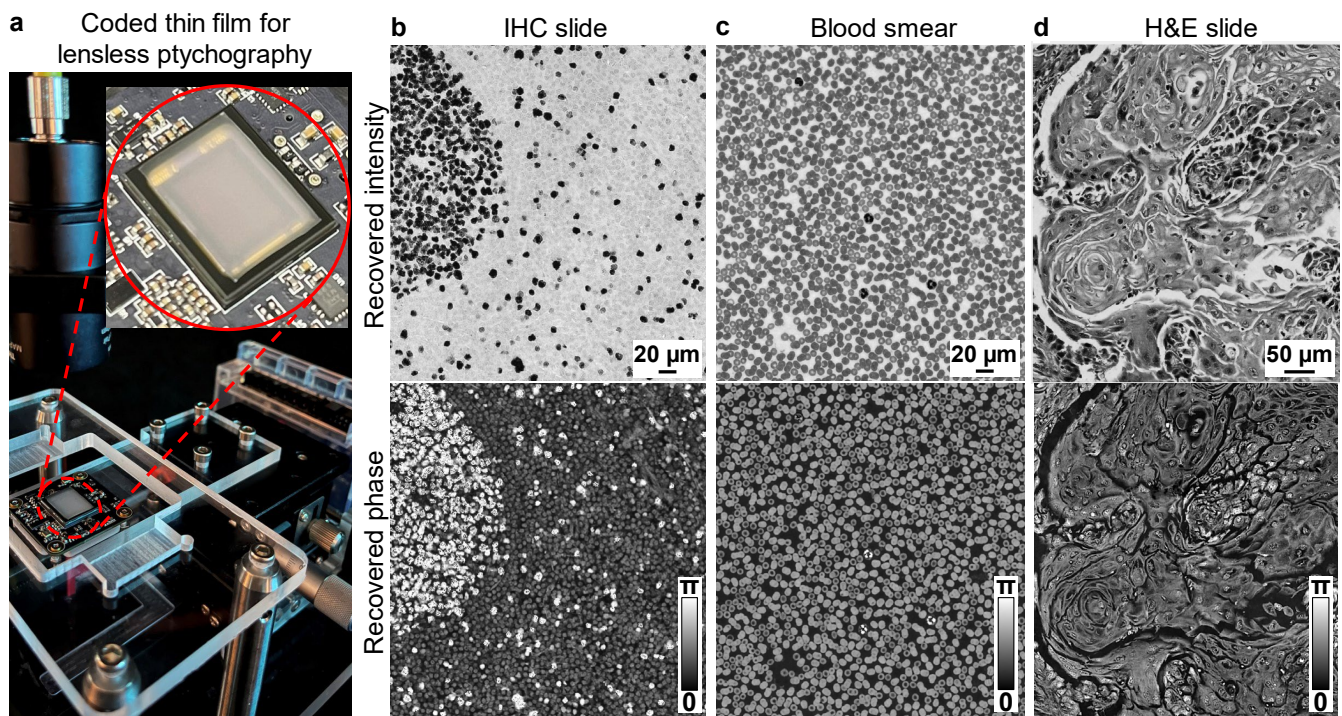

**Figure 8. Utilization of the flexible and detachable coded thin film for lensless ptychography.** (a) The lensless ptychography setup with the coded thin film attached on an image sensor. The recovered intensity and phase images for an IHC slide (b), a blood smear (c), and an H&E slide (d).

Figure 8 illustrates the versatility of the coded thin film, extending its application from lens-based settings to lensless coded ptychography[48-54]. Figure 8(a) showcases the experimental setup where the coded thin film is attached to the coverglass of an image sensor, effectively serving as a computational scattering lens. The use of coded thin film replaces the use of blood cell layers or microbeads as coded surfaces[25,48,51], which previously required permanent attachment to a sensor's coverglass, a process often marred by inconsistencies in uniformity. The reported thin film's flexible and detachable configuration streamlines the operational process for lensless coded detection. During the experiment shown in Figure 8(a), the specimen is placed over the coded sensor, which is then scanned across various lateral positions to capture a sequence of lensless coded diffraction patterns. These patterns are subsequently used for reconstructing the specimens' high-resolution images. Figure 8(b) shows the recovered intensity and phase images of an immunohistochemistry (IHC) slide, delineating details of the cellular structures. Similarly, the blood smear images in Figure 8(c) emphasizes the technique's capability to resolve cells and sub-cellular components, offering a more compact and cost-effective solution for medical diagnostics. Lastly, an hematoxylin and eosin (H&E) stained tissue section in Figure 8(d) further exemplifies the technique's strength in rendering detailed histological features that are vital for pathological examination. The innovative employment of the coded thin film in lensless coded ptychography marks a stride in democratizing high-resolution imaging technology. It holds potential for enhancing point-of-care diagnostics and extending the reach of quality health services to remote locations.

**Discussion and conclusion**
The issue of non-uniform phase transfer in microscopy has historically constrained accurate phase recovery, particularly for low-frequency phase contents. Traditional approaches like the transport-of-intensity equation have attempted to mitigate this by generating intensity contrasts through defocusing, enabling an indirect phase recovery by strategically positioning the specimen out-of-focus[55]. Similarly, matched-illumination methods have tried to enhance phase-to-intensity conversion by aligning the incident illumination angle with the maximum collection angle of the objective lens, particularly at the brightfield-to-darkfield transition zone, to encode phase information into intensity variations[31,56]. While these methods have achieved partial success, they inherently do not provide a uniform phase transfer function across the Fourier spectrum. Simple optical arrangements, such as imaging a linear phase ramp, illustrate the intrinsic limitations of these traditional approaches.

In contrast, the current study demonstrates the potential of scFP to overcome these obstacles by achieving true quantitative phase imaging with a uniform frequency response. Complex wavefront from the object is modulated by the coded thin film layer, resulting in a speckle-like intensity pattern for detection. The validation of scFP across various phase targets, from flat-surfaced crystals to intricate biological samples, confirms its ability to resolve detailed phase information essential for precise scientific measurement and medical diagnostics. Furthermore, the coded thin film's versatility is exemplified in its application to lensless coded ptychography, showcasing a simple strategy that promises to improve point-of-care diagnostics and remote health monitoring through its portability and cost-effectiveness.

For future developments, we aim to extend the application of scFP to Fourier ptychographic diffraction tomography[22,23], wherein the simple addition of the coded thin film is expected to address the common issue of refractive index underestimation. Moreover, the coded thin film introduces a new avenue for measurement diversity in FP. By harnessing both angular and translational diversity, we envision a more robust recovery of three-dimensional specimens, enhancing the detail and accuracy of reconstructions.

In conclusion, the coded thin film's flexible and removable design ensures its seamless integration with a variety of imaging systems, both lens-based and lensless, marking it as a universally applicable tool in diverse imaging scenarios. The advent of scFP is anticipated to ignite novel research avenues and applications, especially in computational imaging, where its accuracy, adaptability, and versatility stand to make a significant impact.

**Acknowledgements**
This work was partially supported by National Institute of Health R01-EB034744 (G. Z.). Kevin Sun acknowledges the support of National Science Foundation high school student research assistantship (NSF MPS-High). Pengming Song acknowledges the support of the Thermo Fisher Scientific fellowship. We thank Dr. Shuhe Zhang at Maastricht University for useful discussions.

**Author contributions**
G. Z. conceived the project. R. W. and P. S. developed the prototype systems and acquired the data. L. Y. and K. S. prepared the coded thin film. R. W., Y. L., and T. W. prepared the display items. All authors contributed to the writing and revision of the manuscript.

**Conflict of Interest**
The authors declare no conflict of interest.

**Data Availability Statement**
The data that support the findings of this study are available from the corresponding author upon reasonable request.